\documentclass[
    ,final            
  ]
  {aipproc}

\layoutstyle{8x11double}

\begin{document}

\title{IPA-CuCl$_3$: a $S=1/2$ Ladder with Ferromagnetic Rungs}


\author{T. Masuda\footnote{Present address: Department of Basic
Science, Yokohama City University, Japan.}~}{ address={Condensed
Matter Sciences Division, Oak Ridge National Laboratory, Oak
Ridge, TN 37831-6393, USA.} }

\author{A. Zheludev}{ address={Condensed Matter Sciences Division,
Oak Ridge National Laboratory, Oak Ridge, TN 37831-6393, USA.} }

\author{H. Manaka}{address={Department of Nano-Structure and
Adv. Materials, Kagoshima University, Kagoshima 890-0065, Japan.}}

\author{J.-H.~Chung}{address={NCNR, National Institute of
Standards and Technology, Gaithersburg, Maryland 20899, USA.}}

\begin{abstract}
The spin gap material IPA-CuCl$_3$ has been extensively studied as
a ferromagnetic-antiferromagnetic bond-alternating $S=1/2$ chain.
This description of the system was derived from structural
considerations and bulk measurements. New inelastic neutron
scattering experiments reveal a totally different picture:
IPA-CuCl$_3$ consists of weakly coupled spin ladders with
antiferromagnetic legs and ferromagnetic rungs. The ladders run
perpendicular to the originally supposed bond-alternating chain
direction. The ferromagnetic rungs make this system equivalent to
a Haldane $S=1$ antiferromagnet. With a gap energy of 1.17(1)~meV,
a zone-boundary energy of 4.1(1)~meV, and almost no magnetic
anisotropy, IPA-CuCl$_3$ may the best Haldane-gap material yet, in
terms of suitability for neutron scattering studies in high
magnetic {f}ields.
\end{abstract}

\maketitle

\begin{figure}
 \includegraphics[width=3in]{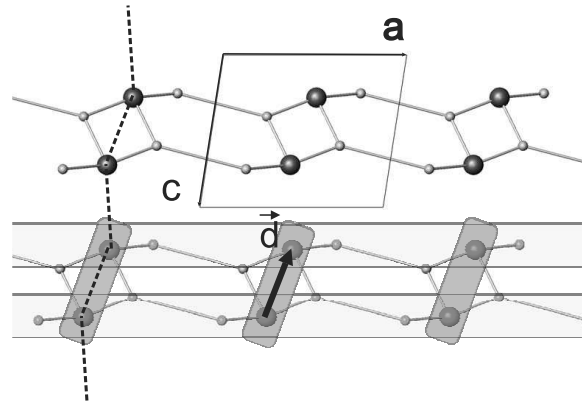}
 \caption{\label{struc} Schematic view of the spin-ladder structure of IPA-CuCl$_3$
 in projection onto the $(a,c)$ crystallographic plane. Large balls are magnetic
 Cu$^{2+}$ ions. Small balls are Cl$^{-}$. The previously proposed alternating-chain model, now shown to be incorrect,
 is represented by a dashed line.}
\end{figure}

Despite years of extensive research, a number of very important
models of one-dimensional (1D) magnets have not yet found their
realizations in real quasi-1D materials. This ``de{f}icit'' became
particularly frustrating when recent attention turned to neutron
studies of high-{f}ield effects in quantum magnets
\cite{Chen2001,ZheludevNDMAP,Stone2003,Ruegg2003,Kenzelmann2004}.
Most known spin-ladder compounds have very large scales of
magnetic interactions and are unsuitable for such studies. A
number of Haldane-gap and related systems with small enough gap
energies have been investigated
\cite{Chen2001,ZheludevNDMAP,Hagiwara2005}, but all feature very
strong single-ion anisotropy that qualitatively alters
high-{f}ield behavior. For this reason, the discovery of the
quasi-1D material (CH$_3$)$_2$CHNH$_3$CuCl$_3$ (IPA-CuCl$_3$) with
almost no anisotropy and a spin gap of $\sim 1.5$~meV was a major
step forward.\cite{Manaka97,Manaka98,Manaka2001} Bulk measurements
on this material revealed a {f}ield-induced Bose condensation of
magnons at a critical {f}ield $H_c\simeq9$~T \cite{Manaka98}. The
original indication was that IPA-CuCl$_3$ should be described in
terms of $S=1/2$ Cu$^{2+}$ -chains running along the $c$ axis of
the triclinic structure (dashed line in {F}ig.~\ref{struc}).
Moreover, it appeared that the exchange interactions connecting
subsequent Cu$^{2+}$ spins alternate between ferromagnetic and
antiferromagnetic ones. Herein we report neutron scattering
studies of the magnetic excitations in IPA-CuCl$_3$ that reveal a
totally different picture.

Fully deuterated IPA-CuCl$_3$ samples were utilized for these
studies. 20 single crystals of total mass 3.5~g were co-aligned to
a cumulative mosaic of 1.5$^\circ$. Magnetic susceptibility and
ESR data are similar to those for the non-deuterated material.
Cell parameters at room temperature are $a=7.766$~\AA,
$b=9.705$~\AA, $c=6.083$~\AA,
$\alpha=97.62^\circ$,$\beta=101.05^\circ$ and
$\gamma=67.28^\circ$, space group $P-1$ \cite{Manaka97}.  The
measurements were done on the SPINS cold neutron 3-axis
spectrometer at NIST, using $E_f=3.7$~meV {f}ixed-{f}inal-energy
neutrons and a BeO {f}{ilter} positioned after the sample. The
scattering plane was either $(h,0,l)$ or $(h,k,0)$. Most data were
taken at $T=1.5$~K.

\begin{figure}
 \includegraphics[width=3.1in]{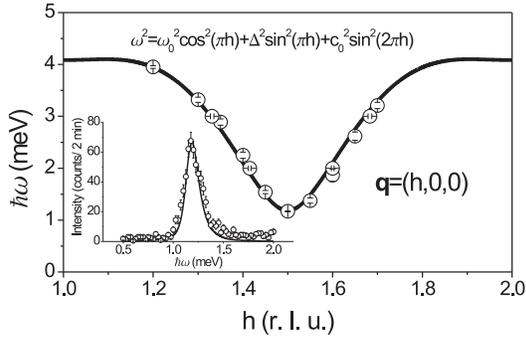}
 \caption{\label{disp} Measured $a$-axis dispersion of the ladder gap excitations in
 IPA-CuCl$_3$. The measured bandwidth along the $c$ crystallographic axis is only about 0.6~meV \protect\cite{Masuda2005}.
 Inset: constant-$q$ scan collected at the 1D antiferromagnetic zone-center $(0.5,0,0)$.}
\end{figure}

From the very {f}irst scans it became obvious that $c^\ast$ is
{\it not} the direction of strong dispersion. This implies that
the bond-alternating $c$-axis chain scenario is {\it not} an
appropriate model for IPA-CuCl$_3$, and that the spin gap must
have a different origin. A systematic survey of the spectrum
enabled us to locate the global energy minimum for the gap
excitations at $\mathbf{q}=(0.5,0,0)$. A typical energy scan at
this wave vector is shown in the inset of {F}ig.~\ref{disp}. The
observed peak position is in good agreement with the gap energy
deduced from bulk measurements.

Additional measurements established that IPA-CuCl$_3$ is indeed
one-dimensional magnet, but with the  $a$-axis being the
strong-coupling direction. The dispersion relation of the gap
excitations along $a^\ast$ was measured in a series of const-$q$
and const-$E$ scans. The resulting dispersion curve is plotted in
{F}ig.~\ref{disp}. These data were analyzed using the empirical
formula shown in the {f}igure\cite{Barnes94}. A good {f}it was
obtained using $\hbar \omega_0=4.1(1)$~meV $\Delta=1.17(1)$~meV
and $c_0=2.2 (1)$~meV (solid line). The energy of the observed gap
mode was also measured as a function of momentum transfer along
the other two reciprocal-space directions. The bandwidth along
$c^{\ast}$ is rather narrow, the gap increasing from its minimum
$\Delta=1.17$~meV at $\mathbf{q}=(0.5,0,0)$ to a maximum value of
$1.8$~meV at $\mathbf{q}=(0.5,0,0.5)$. There is no measurable
$k$-dependence of the gap energy. More details  will be reported
elsewhere \cite{Masuda2005}.

The positions of the magnetic Cu$^{2+}$ ions in IPA-CuCl$_3$ are
non-alternating along the crystallographic $a$ axis. A careful
look at the structure allows to identify uniform {\it spin
ladders} running along this direction, as shown in
{F}ig.~\ref{struc}.  The structural alternation along the $c$ axis
that was the basis of the alternating-chain model ensures a
separation between individual ladders. Along $b$ the ladders are
even better isolated due to non-magnetic layers of the organic
ligand. The ladder model was con{f}irmed by studying the wave
vector dependence of integrated intensities of the gap modes. At
the 1D AF zone-centers $h=0.5$ and $h=1.5$ this intensity is
periodic with $l$ and scales precisely as $\cos^2(\mathbf{qd}/2)$,
where $\mathbf{d}$ is the vector that de{f}ines the ladder rung
(Fig.~\ref{struc}) \cite{Masuda2005}. Such a 3D structure factor
implies that the rung interactions are {\it ferromagnetic}, as
previously proposed \cite{Manaka97}. A ladder of this type can be
viewed as a Haldane spin chain, each pair of rung spins playing
the role of an effective $S=1$. This is especially true for
excitations at wave vectors with $\mathbf{qd}=0$, where the
dynamic structure factor is exactly equal the correlation function
for the total spin on each rung. We conclude that IPA-CuCl$_3$ can
be described as quasi-1D antiferromagnet with ``composite''
Haldane spin chains along $a$.

IPA-CuCl$_3$  turned out not to implement the bond-alternating
chain model as it was assumed to do for quite some time.
Nevertheless, it remains a very interesting and important
low-dimensional material. Future high-{f}ield neutron scattering
studies of this anisotropy-free composite Haldane spin chain
system will be interesting to contrast with recent work on the
strongly anisotropic Haldane gap material NDMAP
\cite{ZheludevNDMAP} and the bond-alternating $S=1$ chain compound
NTENP \cite{Hagiwara2005}.

\begin{theacknowledgments}
  Work at ORNL was carried out under DOE Contract No.
DE-AC05-00OR22725. The work at SPINS was supported by the National
Science Foundation under Agreement No. DMR-9986442.
\end{theacknowledgments}


\end{document}